# On the Shock Source of Sustained Gamma-Ray Emission from the Sun

N Gopalswamy[1], P. Mäkelä[2], S. Yashiro[2], A. Lara[2,3], S. Akiyama[2], and H. Xie[2]

[1]NASA Goddard Space Flight Center, Greenbelt, MD 20771, USA
[2]The Catholic University of America, Washington DC 20064, USA
[3]Instituto de Geofisica, UNAM, Mexico City, Mexico

**Abstract.** It has recently been shown that the spatially and temporally extended γ-ray emission in solar eruptions are caused by >300 MeV protons precipitating on the Sun from shocks driven by coronal mass ejections (CMEs). The γ-rays result from the decay of neutral pions produced in the proton-proton interaction when the >300 MeV protons collide with those in the chromosphere. The evidence comes from the close correlation between the durations of the sustained gamma-ray emission (SGRE) and the associated interplanetary (IP) type II radio bursts. In this paper, we provide further evidence that support the idea that protons accelerated in IP shocks driven by CMEs propagate toward the Sun, precipitate in the chromosphere to produce the observed SGRE. We present the statistical properties of the SGRE events and the associated CMEs, flares, and type II radio bursts. It is found that the SGRE CMEs are similar to those associated with ground level enhancement events. The CME speed is well correlated with the SGRE fluence. High CME speed is an important requirement for the occurrence of SGRE, while the flare size is not. Based on these results, we present a schematic model illustrating the spatially and temporally extended nature of SGRE related to the CME flux rope-shock structure.

## 1. Introduction

Morrison [1] suggested that both continuum and line emissions in the γ-ray band are expected from cosmic sources due to processes such as bremsstrahlung, radiative decay of neutral pions, de-excitation of nuclei, and electron-positron annihilation. Of these he suggested bremsstrahlung continuum and 2.223 MeV line due to neutron capture are likely from the active Sun. Lingenfelter and Ramaty [2] performed detailed calculations of γ-ray emission processes from the Sun. The 2.223 MeV line was first detected by the Gamma-ray Monitor on board the OSO-7 mission. The pion-decay continuum from both the impulsive and late phases was identified by Forrest et al. [3] during the 1982 June 3 γ-ray event using Solar Maximum Mission's Gamma Ray Spectrometer data. A typical γ-ray spectrum showing various lines below ~10 MeV, bremsstrahlung, positronium, and pion continua can be found in [4]. The pion decay continua have three components: γ-rays from the decay of neutral pions, electron and positron bremsstrahlung γ-rays from the decay of charged pions, and positron annihilation continuum from charged pions. Only about a dozen such pion-decay events were recorded by different telescopes between 1982 and 1991 [5]. A few more were added mainly from CORONAS-F/SONG [6]. The primary characteristic of the late-phase γ-ray events is that they last beyond the impulsive phase of the associated flare for minutes to many hours. Of these, only two events had durations >2 h in the pre-Fermi era. Fermi mission's Large Area Telescope (Fermi/LAT [7]) revealed that such late-phase events at energies >100 MeV are rather common [8-10], thanks to the unprecedented sensitivity of the detector in the energy range 20 MeV to 300 GeV. In some cases, the Fermi/LAT events lasted for almost a day [11-13]. These events have been known under various names: Long Duration Gamma-ray Flares (LDGRF [14]), Sustained Gamma-Ray Emission (SGRE [15-16,10,13]) and Late-Phase Gamma-Ray Emission (LPGRE [9]). "SGRE" truly denotes the emission lasting well past the associated flare, so we use it here. The origin of energetic protons responsible for pion production has been controversial for more than 30 years. It was already noted [3] that the protons may be accelerated by a mechanism similar to that responsible for solar energetic particles (SEPs) in space and different from the one responsible for the impulsive particles. Murphy et al. [17] showed indeed that the impulsive phase proton spectrum is too steep (inferred from γ-ray line emission) to produce pion continuum, while the flat, late-phase proton spectrum (producing the pion continuum) is consistent with the hard spectrum of SEPs in the interplanetary (IP) medium. They also suggested that the protons responsible for the late-phase emission are accelerated at coronal shocks. In fact, an intense metric type II burst, indicative of a coronal shock, was observed during the 1982 June 3 event (Solar Geophysical Data, 1982). Akimov et al. [18] agreed with the shock source for the late-phase γ-rays during the 15 June 1991 eruption observed by the GAMMA-1 telescope on board the Gamma Observatory. In a later paper, these authors abandoned the shock source suggesting that the protons from the shock may not precipitate to the Sun from a few solar radii; instead they suggested the protons may be accelerated due to reconnection in the current sheet behind the associated coronal mass ejection (CME) [19]. It has now become clear that such reconnection is long gone well before the end of Fermi/LAT SGRE events [20]. Ryan and Lee [21] suggested that both impulsive and late-phase γ-rays can be explained by energetic particles along with MHD turbulence trapped in flare loops with different turbulence characteristics in the two phases; they had to propose a third phase for SEPs in the IP space, distinct from the late-phase. Kanbach et al. [22], who reported on the longest-duration pre-Fermi-era event of 1991 June 11, also adapted the flare-loop interpretation, but required the loop to be turbulence-free to account for the >8 h duration of the SGRE event.

One of the implications of the flare-loop idea is that the acceleration region is spatially confined to the post-eruption arcade and flare ribbons where the accelerated particles precipitate. The only variability may be the ribbon separation, which is small during the impulsive phase and attains a maximum size in the late phase. The average size of a post eruption arcade is ~90,000 km, while the maximum size is ~200,000 km [23]. The maximum size corresponds to ~17º on the Sun. The spatially-confined nature of the γ-ray source predicted by the flare-loop idea was challenged when 2.223 MeV γ-ray line emission was detected during the 1989 September 29 backside event [24-25]. The active region was ~8º behind the west limb (S25W98). The 2.223 MeV line is produced deep in the chromosphere, so it cannot reach a near-Earth observer from the flare location. The eruption was associated with a fast CME (~1800 km/s) observed by the Coronagraph/Polarimeter on board SMM and a metric type II radio burst. The typical half width of a fast CMEs is ~45º, so there is no difficulty for the particles accelerated in the shock to precipitate on the frontside. Cliver et al. [25] cautioned that the mechanism may or may not apply to the SGRE emission. However, Fermi/LAT observed SGRE emission due to an eruption ~35º behind the east limb on 2014 September 1 [26-30,8,16]. The 45º CME half width is more than adequate for the shock protons to precipitate on the frontside because the shock is typically more extended than the CME flux rope. A 17º-wide post eruption arcade located ~35º behind the limb cannot be the source of γ-rays observed by the Fermi mission.

The spatially and temporally extended nature of the γ-ray source requires an accelerator that is spatially and temporally extended: the CME-driven shock. The solid proof that the SGRE is due to protons from shocks came from the finding that SGRE durations are about the same as the durations of the associated type II radio bursts [13,31], which are excellent indicators of shocks in the corona and IP medium. In this paper, we examine the close connection among SGRE, type II radio bursts, and SEP events, all related to energetic CMEs that seem to be the key solar phenomenon of interest. We also discuss some SGRE events, which seem to be in apparent contradiction to the idea that energetic protons are transported back to the Sun. We use CME observations from the Solar and Heliospheric Observatory (SOHO) and the Solar Terrestrial Relations Observatory (STEREO). The radio data are from the Wind and STEREO spacecraft. We use EUV images from the Solar Dynamics Observatory (SDO) to identify the post eruption arcades in the source regions. We also use GOES and STEREO data on SEP events. Information on the 1991 June 11 SGRE is obtained from published literature and from the Solar Geophysical Data made available by NOAA/NGDC, while the radio data are from the Ulysses mission.

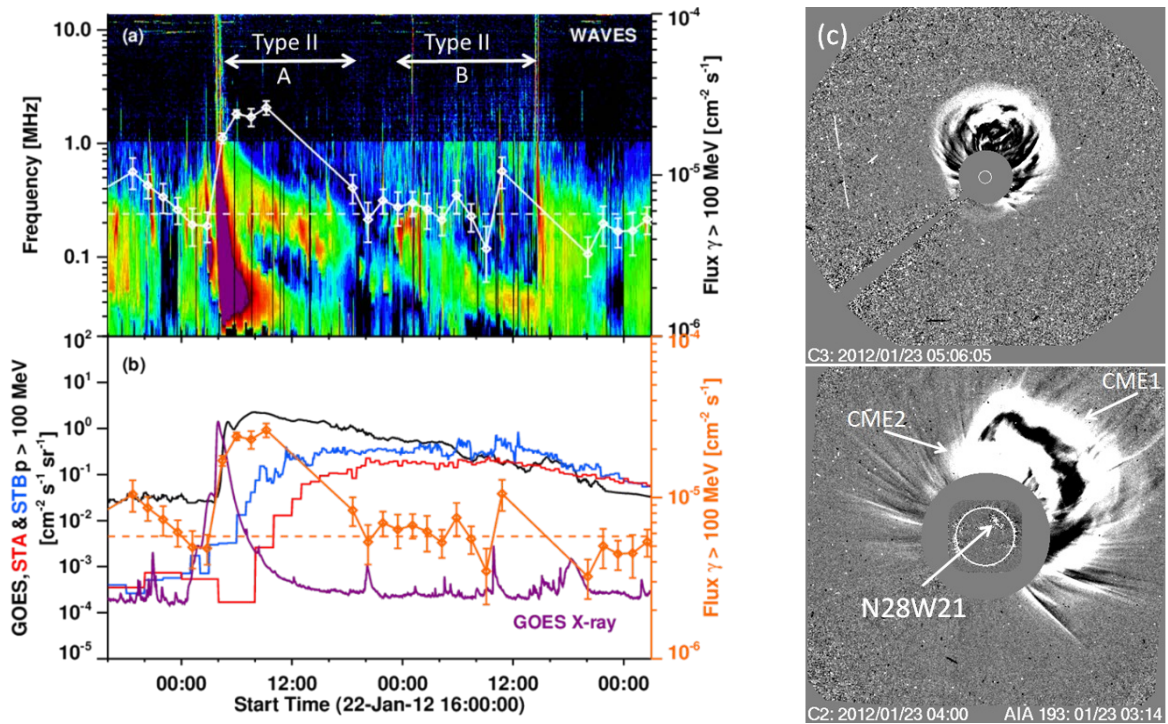

**Figure 1**. (a) The 2012 January 23 SGRE event and the associated type II burst, which has two phases (A, B), separated by ~6 hours. The SGRE duration coincides with the duration of phase A. The vertical feature in the beginning of the type II is a type III burst caused by energetic electrons escaping from the flare site. (b) The SGRE flux plot along GOES soft X-ray light curve and the >100 MeV SEP flux from three spacecraft: GOES, STEREO-Ahead (STA), and STEREO-Behind (STB). The e-folding drop of >100 MeV proton flux roughly occurs at the SGRE end time. (c) The CME responsible for the type II burst from N28W21 at two instances: at 04:00 UT at the onset and at 05:06 UT at the SGRE peak. The two CMEs (CME1, CME2) in the 04:00 UT image merged into a single one at 05:06 UT.

## 2. Fermi/LAT SGRE Events

In order to provide an overview of the SGRE events and the associated phenomena studied in this work, we have shown the 2012 January 23 event associated with a CME from NOAA active region (AR) 11402 located at N28W21. The SGRE is associated with an M8.7 GOES soft X-ray (SXR) flare that starts, peaks and ends at 03:38 UT, 03:59 UT, and 04:34 UT, respectively. High-frequency (e.g., 17 GHz) microwave bursts peak at 03:50 UT, about nine minutes before the SXR peak. This is typical of most flares, so we take the SXR peak time as the end of the impulsive phase and also as the start of the late-phase γ-rays (SGRE). The SGRE peaks around 09:12 UT, about 5 hr after the SXR peak. The eruption location is reasonably connected to Earth, so the >100 MeV proton flux (used as a proxy to the >300 MeV proton flux) sharply increases after the eruption and reaches a broad peak around the time the SGRE attains maximum. Even though STA and STB are not well connected to the eruption, they do observe the SEP event as can be seen from the plots in the lower panel. The end time of the SGRE is determined as the mid time (19:25 UT) between the last signal data point above background (18:35 UT) and the next data point (20:15 UT). Thus the SGRE duration is 15.43±0.83 hr. The CME first appears in the LASCO/C2 field of view (FOV) at 04:00 UT and merges with a preceding CME (CME1) from the same source region. The compound CME has a speed of ~2500 km/s within the LASCO FOV. Consistent with the large source latitude, the central position angle in the early images is ~ 326º but the CME soon becomes an asymmetric halo. The type II burst is observed in the decameter-hectometric (DH) domain starting at 04:00 UT, but a metric type II burst has started at 03:38 UT observed by ground-based radio telescopes. The DH type II has a large break at 19:25 UT, which is also the end time of SGRE. We have labelled this phase of the type II burst as "A" in Figure 1a. After a break of ~6 hr, the type II burst continues until the shock arrival at the Wind spacecraft the next day at 14:33 UT (marked "B" in Figure 1a). Thus, we have taken the end time of the type II between these two times, giving a duration of 24.99±9.57 hr. The SEP event is the second largest (as measured from the >10 MeV flux, 6310 pfu). The e-folding time of the >100 MeV proton flux from the peak also coincides with the SGRE end time. It must be noted that lower energy particles continue to be accelerated until the shock arrives at Earth and even beyond; the particles observed as Earth passes through the shock are the energetic storm particles. The flux of >300 MeV protons greatly diminishes beyond phase A noted above. For the same reason, weak type II bursts continue all the way to shock arrival. The SGRE, flare, CME, SEP, and type II burst characteristics derived this way are listed in Table 1 for 19 Fermi/LAT SGRE events with duration >3 hr (from [13,31]). Also included is the 1991 June 11 pre-Fermi SGRE event that satisfies the duration criterion (>3 h). There was no coronagraph operating during this event, so there is no CME data; however, a CME was observed in the IP medium [31].

Table 1. List of SGRE events with the associated CMEs, flares, and type II bursts

| SGRE | | | CME | | Flare | | SEP | Type II | |
|---|---|---|---|---|---|---|---|---|---|
| Start UT | Dur. (hr) | Fl cm$^{-2}$ | UT | V km/s | Class | Location | pfu | UT | Dur. (hr) |
| 91/06/11 02:09 | 10.38 ±1.5 | 19.5 | ------ | ------ | X12.0 | N31W17 | 1180[c] | 02:01 | 8.67 ±0.67 |
| 11/03/07 20:12 | 21.02±1.55 | 1.08 | 20:00 | 2223H | M3.7 | N31W53 | 50 | 20:00 | 13.17±0.67 |
| 11/06/2 07:46 | 6.98 ±1.47 | 0.12 | 08:12 | 1422H | C3.7 | S19E25 | 0.10 | 08:00 | 4.47 ±0.02 |
| 11/06/7 06:41 | 3.08 ±1.67 | 0.26 | 06:49 | 1321H | M2.5 | S21W54 | 72 | 06:45 | 10.93 ±0.32 |
| **12/01/23 03:59** | 15.43±0.84 | 0.69 | 04:00 | 2511H | M8.7 | N28W21 | 6310 | 03:38 | 24.98 ±9.57 |
| **12/01/27 18:37** | 3.59 ±0.82 | 0.18 | 18:27 | 2541H | X1.7 | N27W71 | 795 | 18:30 | 10.23 ±0.33 |
| **12/03/05 04:09** | 4.25 ±0.80 | 0.06 | 04:00 | 1627H | X1.1 | N17E52 | 100B[b] | 04:00 | 6.99 ±1.08 |
| **12/03/07 00:24** | 20.47±0.84 | 24.1 | 00:24 | 3146H | X5.4 | N17E27 | 6350B | 00:30 | 27.78±6.61 |
| **12/03/09 03:53** | 8.85 ±0.80 | 0.16 | 04:26 | 1229H | M6.3 | N15W03 | 600B | 04:00 | 3.79±1.71 |
| **12/03/10 17:44** | 11.62 ±1.59 | 0.05 | 18:00 | 1638H | M8.4 | N17W24 | 120B | 17:55 | 7.42±0.42 |
| 12/05/17 01:47 | 3.08 ±0.80 | 0.10 | 01:48 | 1596H | M5.1 | N11W76 | 255[c] | 01:40 | 4.73 ±0.20 |
| **13/05/13 02:17** | 7.47 ±1.62 | 0.13 | 02:00 | 1270H | X1.7 | N11E90 | 25[b] | 02:20 | 6.13 ±0.28 |
| **13/05/13 16:05** | 8.69 ±0.95 | 0.40 | 16:07 | 1852H | X2.8 | N11E85 | 400B[b] | 16:15 | 8.58 ±0.17 |
| **13/05/14 01:11** | 5.99 ±0.82 | 0.35 | 01:25 | 2645H | X3.2 | N08E77 | 608B[b] | 01:16 | 6.23 ±0.83 |
| **13/05/15 01:48** | 3.61 ±0.81 | 0.03 | 01:48 | 1408H | X1.2 | N12E64 | 42[b] | 01:49 | 6.41 ±0.28 |
| 14/02/25 00:49 | 8.46 ±1.59 | 14.9 | 01:25 | 2153H | X4.9 | S12E82 | 400[b] | 00:56 | 10.32 ±0.75 |
| 14/09/01 11:11 | 3.92 ±0.76 | 7.94 | 11:12 | 2017H | X2.4[a] | N14E127 | 3083[b] | 11:12 | 7.5 ±2.25 |
| 15/06/21 02:36 | 14.06 ±1.61 | 27.1 | 02:36 | 1740H | M2.6 | N12E16 | 1066 | 02:33 | 20.95 ±1.50 |
| **17/09/06 12:02** | 18.43 ±0.80 | 1.01 | 12:24 | 1884H | X9.3 | S08W33 | 844B | 12:05 | 20.92 ±1.00 |
| **17/09/10 16:06** | 15.18 ±0.81 | 13.6 | 16:00 | 3165H | X8.2 | S09W92 | 1490B[c] | 16:02 | 15.72 ±0.92 |

[a]estimated; [b]from STEREO-B; B – the SEP event had high background; [c]GLE;1 pfu = 1 particle/cm$^2$.s.sr

For each SGRE event, the duration and fluence (Fl) are listed in columns 2 and 3. The CME speeds are three-dimensional speeds obtained from the graduated cylindrical cell (GCS [33]) model that fits a flux rope and a shock to the CMEs. The CMEs are observed by SOHO's Large Angle and Spectrometric Coronagraph (LASCO [32]) and STEREO's Sun Earth Connection Coronal and Heliospheric Investigation (SECCHI [34]) observations. The suffix 'H' to the speed values indicates that the CME is a full halo. The listed soft X-ray flare sizes and the heliographic coordinates of the eruption location are from NOAA's Space Weather Prediction Center. The SEP intensity listed is the maximum value in pfu attained by >10 MeV protons. These values are obtained from GOES data compiled at the CDAW Data Center (https://cdaw.gsfc.nasa.gov/CME_list/sepe/). For eastern events, the SEP intensity is computed from the STEREO high-energy telescope (HET) and low-energy telescope (LET) data. The equivalent >10 MeV intensity is obtained by first computing SEP spectra at energies below 100 MeV and extrapolating to higher energies (see [35] for details). The type II burst start time and duration are from the Radio and Plasma Wave (WAVES [36]) experiment on Wind compiled at the CDAW Data Center (https://cdaw.gsfc.nasa.gov/CME_list/radio/waves_type2.html), except that the 2014 September 1 event data are from STEREO/WAVES. Table 1 is the basis for many plots and results presented in this paper.

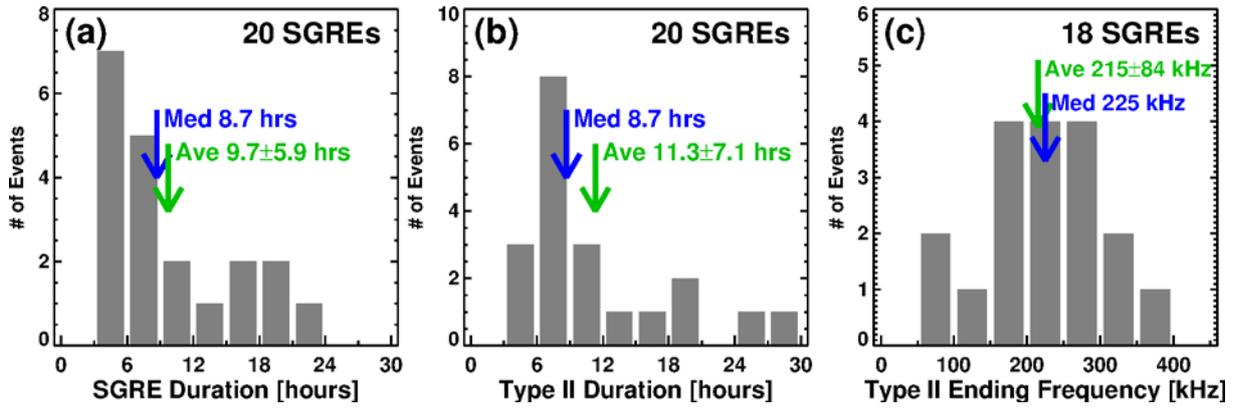

**Figure 2**. Distributions of SGRE durations (left), type II burst durations (middle) and ending frequencies of type II bursts (right). The median (Med) and average (Ave) values are noted on the plots.

### 3. SGRE and type II burst durations

The close relation between SGRE and type II bursts is evidenced by their durations. The type II burst duration is closely related to the burst's ending frequency. Since the type II burst is emitted at the local plasma frequency or its harmonic in the vicinity of the shock, the ending frequency corresponds the distance from the Sun where such plasma frequencies (and hence electron densities) prevail. Figure 2 shows distributions of the SGRE durations, type II durations, and ending frequencies for the events in Table 1. The range and average durations of the two electromagnetic emissions are similar. The average duration of type II bursts is about 1.5 hr longer mainly because of the two intense events that had a break in the dynamic spectrum but the type II burst continued intermittently until the shock arrived at the observing spacecraft (one of them is in Fig. 1 and the other is the 2012 March 7 event). In these two cases, the error bars were very generous resulting in the higher average duration.

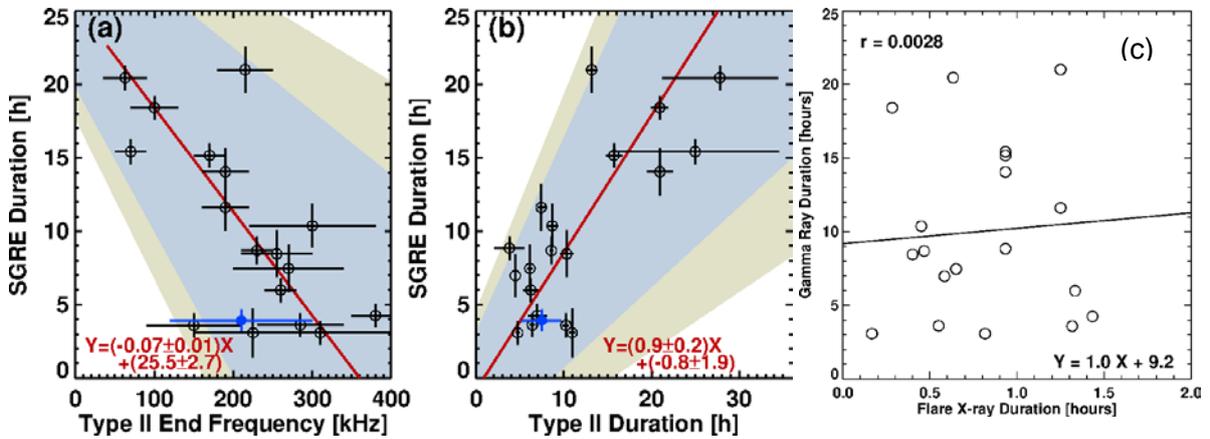

**Figure 3.** The scatter plot of SGRE durations with the ending frequencies (a) and durations (b) of type II bursts. The blue data points correspond to the 2014 September 1 backside event (not included in the correlation). The blue and yellow shaded areas denote the 95% and 99% confidence levels. (c) The scatter plot between SGRE duration and the SXR flare duration showing no correlation.

The close relationship of SGRE durations with type II ending frequencies and durations found for SGRE events with >5 h duration [13] also holds when we included events with >3 durations. Figure 3 shows the corresponding scatter plots. The correlations indicate a clear linear relationship: for longer duration SGRE events, the underlying shock travels a larger distance from the Sun, and the SGRE ends when type II burst ends. This result is remarkable because the energetic particles responsible for the two types of emissions originate from the same shock, but the electromagnetic emissions are produced at different locations: type II bursts in the vicinity of the shock and SGRE at the Sun where the >300 MeV protons precipitate after traversing the Sun-shock distance. On the other hand, the SXR flare duration has no correlation with the SGRE duration.

The SGRE scenario inferred from the SGRE – Type II burst correlations is illustrated in Fig. 4. Protons diffusing back to the Sun along open magnetic field lines threading the shock can produce both neutral and charged pions. At photon energies >100 MeV, the main contribution comes from the decay of neutral pions into γ-rays. Charged pion decay results in electron and positron bremsstrahlung as well as positron annihilation γ-rays, but the flux of these photons is more than an order of magnitude weaker [17]. There may also be a small contribution to the electron bremsstrahlung from high-energy electrons diffusing back to the Sun from the shock. One of the implication of the model in Fig. 4 is the spatially extended nature of the γ-ray source. The spatial extent is determined by the extent of the nose region of the shock where the highest energy particles are accelerated [37] and the field lines that are accessed by these particles. The flux rope – shock structure requires that particles precipitate along open field lines that lie on the outskirts of the flux rope because only these field lines connect to the shock nose. This suggests that the largest spatial extent of the source should be in the direction connecting the foot-points of the flux rope. For example, for a high-inclination flux rope, the γ-ray source is extended in the north-south direction; for a low-inclination flux rope, it is extended east-west [38]. Figure 4b shows the post eruption arcade and the core dimmings during the 2015 June 21 SGRE indicating a high-inclination flux rope (see [39]). Spatial localization of SGRE from this event needs to be done to verify the asymmetry in spatial extension. Another important point is the field strength in the core dimming region is close to the quiet Sun values, not the active region values [39].

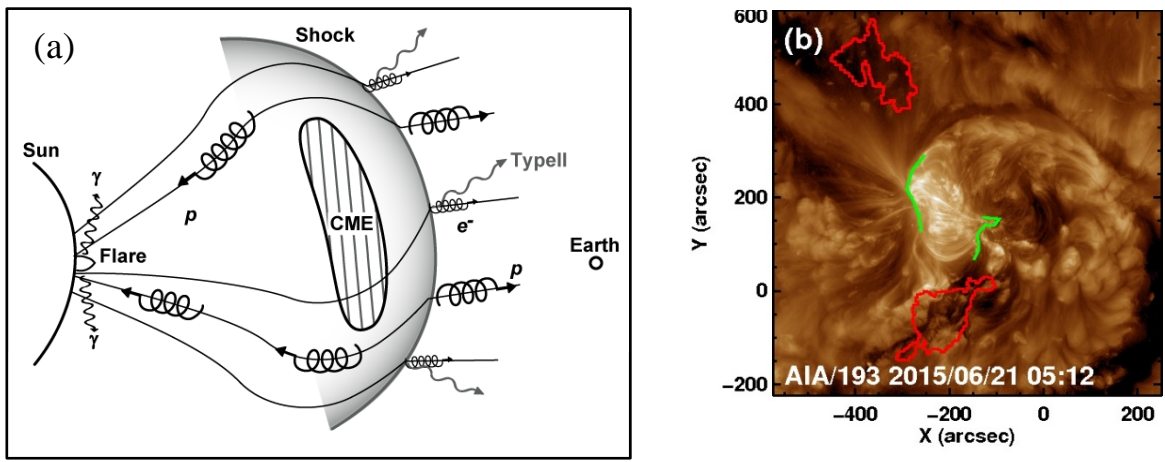

**Figure 4**. (a) Schematic model showing the propagation of protons (p) and electrons (e⁻) from the CME-driven shock toward and away from the Sun. Low-energy electrons (~10 keV) produce type II bursts in the vicinity of the shock. Energetic protons and electrons can be observed as SEP events when the observer is connected to the field line on which these particles propagate. A representative flare loop is shown at the Sun. (b) Two types of magnetic structures during the eruption associated with the 2015 June 21 SGRE event: the post eruption arcade (feet marked in green) and the CME flux rope (feet marked by the red contours). The feet of the flux rope correspond to the core dimmings in the 193 Å EUV image obtained by SDO's Atmospheric Imaging Assembly (AIA).

Figure 5 shows the CME associated with the 2014 February 24 SGRE event as observed by STEREO coronagraph COR2 and Heliospheric Imager (HI). When the SGRE ended, the shock is at a distance of ~70 Rs; the CME also slows down to ~1200 km/s. Thus, the source of protons is also located up to tens of solar radii by the time SGRE ends. In another well-observed case (2015 June 21), the shock distance was obtained as ~90 Rs from the drift rate of the associated type II burst [13,39]. Such distances are consistent with the ending frequency of type II bursts, which is the local plasma frequency at such distances (see Fig. 3).

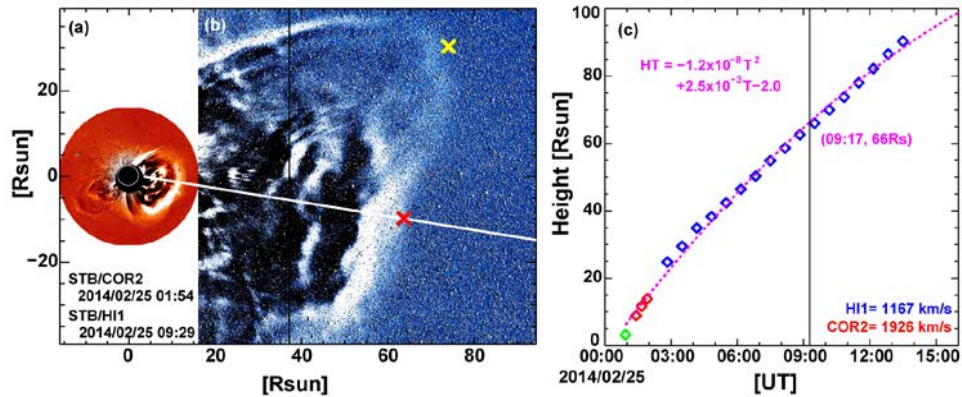

**Figure 5**. The 2014 February 25 CME in the COR2 FOV (a), in the HI-1 FOV (b) and the height (HT) -time (T) plot (c) measured along the white line in (b). Images (a) and (b) correspond to SGRE peak and end. When SGRE ended, the CME leading edge was at ~66 Rs from the Sun's center. The yellow cross marks probably the shock location around 70 Rs.

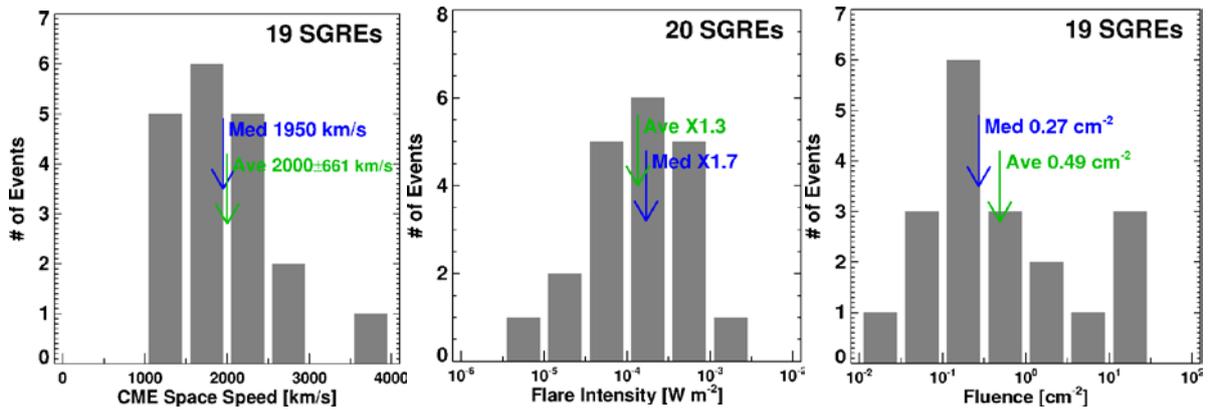

**Figure 6.** Distribution of deprojected (space) CME speeds (left), SXR flare sizes (middle) and fluence (right) of SGRE events in Table 1. The median (Med) and average (Ave) values are noted.

## 4. CMEs and flares associated with SGRE events

The CME speeds listed in Table 1 are averaged over the coronagraph FOV. The sky-plane speed measured from LASCO alone can also be deprojected using a cone model and the results are not too different (see https://cdaw.gsfc.nasa.gov/CME_list/halo/halo.html). We see that SGRE CMEs are all very fast (column 5 in Table 1 and Fig. 6), with an average speed of 2000±661 km/s. Such high speeds (10.3 Rs/hr), when combined with the average SGRE duration (9.7 hr), gives a shock distance of ~100 Rs (see Fig. 5) by the time the SGRE ends. The SGRE CME speed is similar to the average speed of CMEs responsible for ground level enhancement (GLE) in SEP events [40]. In addition, all these CMEs are full halo CMEs as indicated by the suffix "H" to the speeds in Table 1. Halo CMEs are known to be inherently more energetic compared to the general population [41-43]. The larger the halo fraction the higher is the average kinetic energy of a CME population. All SGRE CMEs being halos represents the highest halo fraction even higher than that in GLE CMEs. Shocks of such CMEs are known to accelerate particles to energies >1 GeV, so the presence of >300 MeV protons required for producing the pion-continuum is guaranteed. It must be noted that both the GLEs of solar cycle 24 (2012 May 17 and 2017 September 10) are SGRE events listed in Table 1.

The SXR flare sizes of the SGRE events (see Table 1 and Fig. 6, middle) are generally large: 12 X, 7 M, and 1 C. While it is known that faster CMEs are generally associated with larger flares, it is remarkable that ~42% of the flares are not X-class flares. However, as noted above, all the events were associated with fast halo CMEs. Such a distribution of flare sizes was also noted in GLE events [40]. We computed the γ-ray fluences from the light curves considering 5-min bins by extrapolating between Fermi/LAT data points. The γ-ray fluence varies over three orders of magnitude with an average value of ~0.5 cm$^{-2}$ (see Fig. 6, right). The scatter plots of SGRE fluence with CME speed and SXR flare fluence are reasonably correlated with similar correlation coefficients (Fig. 7). These correlations are statistically significant as the correlation coefficients exceeded the Pearson's critical correlation coefficient. The probability (p) that such high correlations are obtained by chance is 0.005.

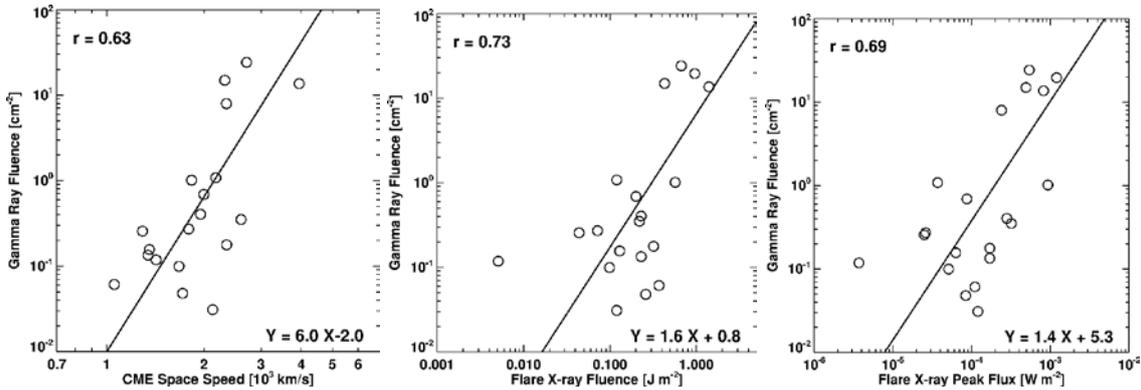

**Figure 7.** The scatter plot of SGRE fluence with CME speed (left), SXR flare fluence (middle), and SXR peak flux (right). The fluence is calculated by interpolating linearly between >100 MeV γ-ray flux data points. In the regression equations, X denotes (in log units) CME speed in 10$^3$ km/s (left), SXR fluence in J m$^{-2}$ (middle) and SXR peak flux in W m$^{-2}$ (right); Y is the gamma ray fluence in cm$^{-2}$ in log units. All correlations are statistically significant: The Pearson critical correlation coefficient is 0.575 (p=0.005 for 19 CMEs and peak SXR fluxes) and 0.590 (p =0.005 for 18 flare fluences). The correlations are slightly different since the 1991 June 11 event is not included in the speed plot (no CME data), while the SXR fluence and peak flux are available from Solar Geophysical Data.

**5. SGRE CME source locations**
The source locations of SGRE eruptions are all in the active region belt (see Table 1, column 7), where the magnetic field strengths are known to be high and required to power the energetic CMEs. The latitudes are higher early in the solar cycle (e.g., events in 2011) and become lower as the cycle progresses (e.g., events in 2013 onward) reflecting the sunspot butterfly diagram. The sources are at all longitudes, with no preference for the eastern or western hemisphere. This is contradictory to the report by Winter et al. [44], who claimed that SGRE sources occur preferentially from the north-eastern quadrant of the Sun. In fact, the SGRE events during March 5-10, 2012 have source locations ranging from E52 to W24. Similarly, the source regions of the 1991 June 4-15 SGRE events in the pre-Fermi era started in the eastern hemisphere (E70 on June 4) and ended in the western hemisphere (W69 on June 15, see [45] for the evolution of the source active region 6659). The June 4-9 SGRE events occurred in the eastern hemisphere, while the intense ones on June 11 and 15 occurred in the western hemisphere. Among the 38 Fermi/LAT SGRE events listed in [44], 34 have known source locations, which are evenly split between the eastern and western hemispheres, just like our events in Table 1. Even the X-class flares without SGREs in [44] are evenly distributed between the eastern and western hemispheres. We do see more source locations in the northern hemisphere (15 out of 20 events in Table 1). This is due to the double-peak nature of the sunspot cycle 24 in which the northern (southern) hemisphere had more active regions during the first (second) peak [46].

**6. SEP events and SGRE**
Table 1 shows that all SGRE events are associated with large SEP events either at Earth or at STEREO-Behind (STB) as indicated by the >10 MeV proton intensity in particle flux units (pfu). STB is well connected to SGRE events originating from the eastern hemisphere in Earth view. There is one exception, the 2011 June 2 event that has no significant SEPs, although there is a type II burst indicative of particle acceleration. This event will be discussed later. Many SEP events occur when the background is high due to preceding events. If a shock has access to the background particles, one expects an efficient acceleration of these particles. It has been noted that only about half of the SEP events are associated with SGRE [47]. Not all large SEP events are expected to have SGRE because some have soft spectra. The soft spectrum can result when the underlying CME forms shocks at large distances from the Sun [35,48], or when the shock nose is not connected to the observer [37]. SEPs from CMEs associated with filament eruptions outside active regions have a soft spectrum because of the low acceleration efficiency at larger distance from the Sun due to the decreasing magnetic field. The shock is the strongest at the nose, so the highest-energy particles are accelerated there. On the other hand, lower energy particles can be accelerated over the whole shock surface. The SEP event associated with the 2011 March 7 SGRE has a soft spectrum (fluence spectral index is 4.70 compared to 2.68 for GLE events, see [35]). This event has been considered as a counter example for the SEP cause of SGRE because it lacks high-energy SEPs. Figure 8 shows the GOES proton intensity plots at various energy channels and a snapshot of the associated CME heading in the north-west direction at a central position angle of 313⁰. The direction is consistent with the far-north source location (N31W53) and the solar B0 angle (-7⁰.25). This means the shock nose is not connected to GOES 13 in the ecliptic and hence the high-energy particles are not expected to be detected, although they precipitate at the Sun and produce SGRE. The shock flank, where only low-energy particles are accelerated, is connected to GOES resulting in the observed soft spectrum. The situation is similar in the case of the 2012 January 23 event, but the spectrum is slightly harder. This is likely because the nose extent of this CME is larger than that of the 2011 March 7 CME.

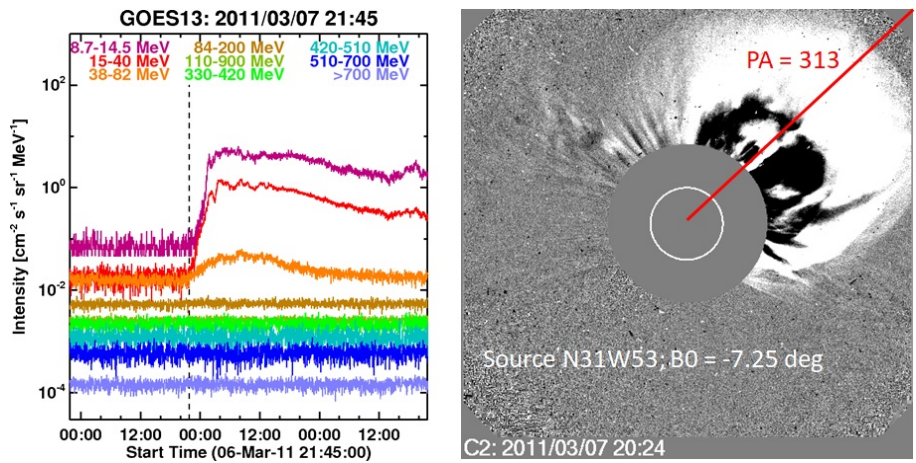

**Figure 8.** (left) GOES 13 SEP data showing no particles in energy channels higher than 38-82 MeV. (right) The underlying CME heading along position angle (PA) 313⁰, which is 43⁰ above the ecliptic. Given the source location at N31W53 and a solar B0 angle of -7⁰.25, it is clear that the shock nose is ~38⁰ from the ecliptic, which too far for the high-energy particles to arrive at Earth.

## 7. SGRE clustering and preconditioning of the ambient medium

In column 1 of Table 1 we have listed several SGRE events in bold font. The events occurred in clusters (originating from the same active region). There are 4 clusters involving 12 SGRE events (or 63%) among the 19 Fermi/LAT events. The 2012 March events and 2013 May events form the two largest clusters (4 SGREs in each). The longitudinal range over which the CMEs erupted is 38⁰ and 36⁰ for these two clusters. Note that these values are typically less than the half width of energetic CMEs. The 2012 January 23 and 27 events from the same active region are separated by a longitudinal extent of ~50⁰, while the September 2017 events are separated by ~59⁰. Even the 1991 June 4-15 events [49] occurred in a cluster of 5, two of them lasting for >2 h (June 11 and 15). The clustering implies that the ambient medium is highly disturbed and filled with energetic particle populations.

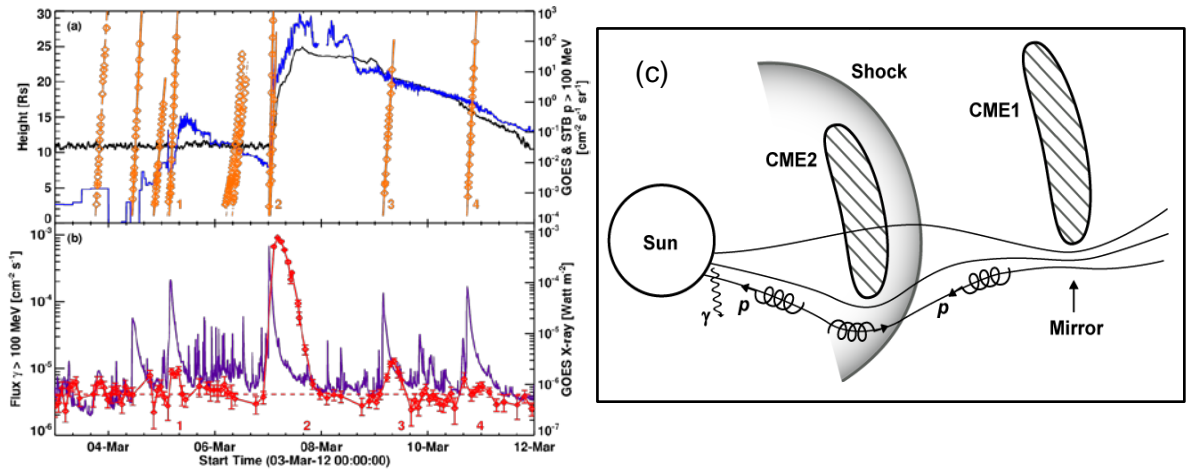

**Figure 9.** One of the SGRE clusters with 4 events (numbered 1-4): (a) CME height-time plots (orange) over-plotted on >100 MeV proton flux curves (blue –STB, black – GOES) and (b) >100 MeV SGRE flux and SXR flux showing the flares. (c) A scenario in which a magnetic mirror is formed near CME1 that reflects most of the particles from CME2's shock back to the Sun.

Figure 9a,b show the SGREs, CMEs, SEPs, and flares during the 2012 March cluster. In addition to the CMEs that caused the SGRE events, other CMEs from the same active region are also shown in the height-time plots. The >100 MeV proton flux is also shown over the period of the SGRE cluster. It is clear that the >100 MeV proton flux is the lowest for SGRE1 and the highest for SGRE2. The proton flux is declining after SGRE2, but boosted slightly higher due to CME3 and CME4 responsible for SGRE3 and SGRE4, respectively. The CME of SGRE1 is preceded 3 CMEs, while that of SGRE2 is preceded by 2 CMEs from the same source region. SGRE2 is actually associated with two CMEs in quick succession, the first one being faster (3146 km/s) than the second (2160 km/s) as can be seen from the slightly different slopes of the closely-spaced height-time plots. Both CMEs are shock driving, but there is no chance of the second CME overtaking the first one. Such a combination seems to have resulted in a copious supply of >300 MeV protons precipitating at the Sun: the first CME helps reflect particles accelerated by the second CME. The 2012 March cluster shows later CMEs in the cluster are launched into a medium highly disturbed by preceding CMEs, energetic particles, and possibly enhanced turbulence. Such a situation is termed as preconditioning and thought to influence the intensity of SEP events [50-51]. The ambient medium may take about 2-5 days to relax back after the passage of a large CME [52], so the preconditioning is likely to be a source of variability in SGRE events. The type of effect one expects depends on the relative timing, kinematics, and distance between the primary and preceding CMEs.

Now we consider the 2011 June 02 SGRE event, which has a duration >6 hr, but there is no associated SEP event. The CME is associated with a DH type II radio burst, indicative of particle acceleration by the shock. The speed (~1100 km/s) of the underlying CME at 08:12 UT is well below the average (~2000 km/s) but similar to the average speed of CMEs associated with SEP events and DH type II bursts. The source, located at S19E25, is not well connected to Earth. For STB (located at E93), the source is at W68 and hence well connected, but the >10 MeV SEP flux is negligible at STB also. This can happen if particles accelerated at the shock of the primary CME are reflected back to the Sun by the mirror formed near the preceding CME at 18:26 UT on June 01. The expected scenario is shown in Fig. 9c, which can be thought of as an extreme case in which the preceding CME does not allow any SEPs to go past the preceding CME to be detected by the observer.

While the mirror scenario in Fig. 9c explains the lack of an SEP event, there is another mirroring issue often brought in against the shock scenario [28]: the strong mirror force may prevent protons from the shock source from precipitation. However, this is a common problem for closed loops in post eruption arcades and open filed lines in the periphery of the CME flux ropes. The mirror ratio is likely to be the same in both cases: flare loops have high field strengths at the reconnection region and foot points, while the open field lines have weaker field both at the shock and at the quiet chromosphere in the flux rope foot pints located in the outskirts of the active region (see Fig. 4b).

**8. Summary and conclusions**

We investigated how the >100 MeV SGRE events with duration >3 hr are related to IP type II radio bursts, large SEP events, and white-light CMEs in order to understand the source of >300 MeV protons responsible for the observed γ-ray continuum. We developed a schematic model based on the observational result that the SGRE ends around the same time the type II burst ends. The ultimate energy source is the magnetic energy release at the Sun in the form of fast and wide CMEs that drive a shock in the corona and IP medium. The same shock accelerates ~10 keV electrons that produce type II radio bursts and >300 MeV protons responsible for the observed γ-rays. Although the precipitating protons produce charged and neutral pions, at photon energy >100 MeV, the dominant contribution is from the decay of neutral pions. Comparison of the radio dynamic spectrum with the CME imagery indicates that the CME-driven shock may be located roughly half way between the Sun and Earth when SGRE ends in many cases. The ending frequencies of the type II bursts are consistent with the local plasma frequency prevailing at such distances. CMEs with such low ending frequencies for type II bursts are known to be very energetic, consistent with 100% halo fraction and ~2000 km/s average speed of the SGRE CMEs. Another strong support for the proposed shock scenario is that the SGRE CMEs have the same properties as GLE CMEs confirming the acceleration of GeV particles and hence >300 MeV particles. The SGRE CME speed and SXR fluence are well correlated with the SGRE fluence, suggesting major energy release in SGRE source regions. The majority of SGRE events occurred in clusters suggesting the availability of seed particles for later shocks in a given cluster.

The association between SEP and SGRE events is less than perfect because of the magnetic connectivity and the spectral hardness requirements. Inherently soft-spectrum events are not expected to produce SGREs because they lack the required high-energy particles (e.g., the SEP events associated with CMEs from filament eruption regions outside active regions). Poorly connected SEP events may still produce SGREs, but the particles are not detected by the observer. The poor cadence of Fermi/LAT may miss some SGRE events even though there are large SEP events with >300 MeV protons [9]. Finally, the exceptional case of an SGRE event with no SEP event can be explained by a propagation issue in that a magnetic mirror formed near a preceding CME reflects particles back to the Sun to produce γ-rays, but no particles escape to be detected as SEP events.

**Acknowledgments**
We benefited from the open data policy of Fermi, SOHO, Wind, STEREO, SDO, and Wind missions. This work was supported by NASA's Living With a Star program. PM was partially supported by NASA grant NNX15AB77G. HX was partially supported by NASA grant NNX15AB70G.